\documentclass[%
 reprint,
 amsmath,amssymb,
 aps,
]{revtex4-2}
\usepackage[pdftex, pdftitle={Article}, pdfauthor={Author}, bookmarks=true, colorlinks, linkcolor=blue,
urlcolor=blue, citecolor=blue, plainpages=false, pdfpagelabels,
final, breaklinks=true]{hyperref}
\usepackage{graphicx}
\usepackage{dcolumn}
\usepackage{bm}
\usepackage{booktabs} 
\usepackage{multirow} 

\begin{document}

\title{Continuously controllable dissipative and coherent couplings by the interaction between anti-resonance and multiple magnons}

\author{Zhenhui Hao, Yuping Yao, Kang An, Xiling Li}
\author{Chi Zhang} 
    \email[Correspondence email address: ]{zc@lzu.edu.cn}
\author{Guozhi Chai}
    \email[Correspondence email address: ]{chaigzh@lzu.edu.cn}
\affiliation{Key Laboratory for Magnetism and Magnetic Materials of the Ministry of Education, Lanzhou University, Lanzhou, 730000, People's Republic of China.}

\date{\today} 

\begin{abstract}
We experimentally realize the continuously controllable dissipative coupling and coherent coupling induced by different magnon modes and the same anti-resonance.
It has been observed that the weaker the microwave magnetic field distribution of the magnon mode in magnetic materials, the more likely dissipative coupling is to occur.
Conversely, stronger magnetic field distributions favor coherent coupling.
Based on this principle, we have designed and implemented a system that alternates between dissipative and coherent coupling regimes.
 It allows microwave signals to be selectively transmitted over a large applied magnetic field range at the frequency of anti-resonance.
Our experimental achievements may promote the construction of new magnonics devices like magnetic-tuning switch.
\end{abstract}

\maketitle

\emph{Introduction.}--Cavity magnonics, founded on the interactions between magnons (quantized excitation of spin wave\cite{Nat.Phys.11.453}) and cavity photons (the quanta of electromagnetic waves\cite{Nature.118.874}), has emerged as a potent platform for investigating hybrid quantum systems\cite{sciadv.1501286,SolidStatePhys.69.47,SolidStatePhys.70.1,SolidStatePhys.71.117}. 
Magnon is crucial in examining the dynamic characteristics and quantum processes of magnetic materials, including low energy consumption\cite{J.Appl.Phys.128.161101}, nanometer-scale wavelengths\cite{Nat.Commun.11.1445,Nat.Commun.9.738,Nat.Nano.11.948-953,Appl.Phys.Lett.99.162502}, nonlinear phenomena\cite{Sci.Adv.9.4609,Phys.Rev.X.9.041036}, and varied dispersion relationships\cite{Sci.366.6469,Sci.367.6476,Nat.Nano.18.1000–1004}, which establish the core status in magnetism and come into notice by researchers.  
In this framework, information is carried and transmitted via polaritons produced through the coherent coupling between magnons and photons\cite{PhysRevLett.111.127003,PhysRevLett.132.156901,Appl.Phys.Lett.115.022407,J.Phys.D:Appl.Phys.52.305003,Appl.Phys.Lett.119.132403,arXiv:2304.09627}.
This new horizon have significantly impacted quantum informatics\cite{Sci.Adv.3.e1603150}, spintronics\cite{PhysRevLett.114.227201,PhysRevLett.118.217201}, cavity optomagnonics\cite{PhysRevLett.116.223601,Annalen.der.Physik,PhysRevLett.118.107205,PhysRevLett.117.123605}, and quantum magnetism\cite{science.aaa3693}. 
In addition, a kind of distinctive coupling mode defined as dissipative magnon-photon coupling was identified\cite{J.Appl.Phys.127.130901,J.Appl.Phys.129.201101}, which has since been swiftly corroborated across diverse systems with varying cavity configurations, including waveguide\cite{PhysRevLett.121.137203,PhysRevB.100.214426}, three-dimensional resonators\cite{New.J.Phys.21.065001,New.J.Phys.21.125001} and planar resonators\cite{PhysRevB.99.134426,PhysRevApplied.11.054023,PhysRevLett.123.127202,PhysRevB.101.064404,PhysRevApplied.18.044074}.
Notably, due to the dissipative coupling in these open\cite{PhysRevB.99.134426,PhysRevApplied.11.054023} or quasi-closed systems\cite{PhysRevLett.121.137203,New.J.Phys.21.065001,PhysRevB.100.094415}, loss is transformed from a hindrance into a beneficial mechanism for system control, facilitating novel applications. Nonreciprocal coupling can be achieved and controlled  by dissipative coupling and coherent coupling synergistically\cite{PhysRevLett.123.127202,PhysRevLett.125.147202}, paving the way for advancements in cavity magnonics to create devices with distinctive functionalities, including non-reciprocal wave propagation\cite{PhysRevB.103.184427,arXiv:2405.17869} and long-distance transmission\cite{PhysRevLett.131.106702,PhysRevLett.132.206902,Nat.Commun.8.604,PhysRevX.5.021025,PhysRevApplied.7.024028}.

Previous researches prefer the couplings induced by ferromagnetic resonance (FMR) mode and cavity mode,  and the coherent-dominated coupling and dissipative-dominated coupling do not occur  simultaneously, but only selectively\cite{sciadv.1501286,SolidStatePhys.69.47,SolidStatePhys.70.1,SolidStatePhys.71.117}.
The utilization of the same cavity photon mode to achieve coherent and dissipative couplings synchronously within a system has emerged as a significant challenge in the advancement of cavity magnonics.
In previous studies, we achive the coherent multimode  couplings induced by cavity mode, FMR mode and forward volume magnetostatic spin waves (FVMSW, a kind of magnon mode) modes\cite{PhysRevB.103.184427}; Rao $et \ al.$ introduced the dissipative coupling of FMR mode and anti-resonance mode\cite{New.J.Phys.21.065001}.
Subsequently, Castel $et \ al.$ provided a comprehensive analytical description of the dissipative coupling associated with an anti-resonance within a hybrid system comprising a quasi-closed cavity\cite{PhysRevApplied.22.064036}. 

According with these researches, we  realize the simultaneous coherent-dominated coupling and dissipative-dominated coupling (strictly speaking, coherent and 
 dissipative coupling mentioned later is called coherent-dominated and dissipative-dominated coupling) induced by an anti-resonance mode and multiple magnon modes in the quasi-closed cavity with an yttrium iron garnet (YIG) wafer.
We observe that the weaker the microwave magnetic field distribution of magnon modes at the resonance frequency, the more likely dissipative couplings are to occur; conversely, coherent couplings happen. 
Our results provide a priori reference for exploring the hybrid cavity magnonic system and put forward a new type of magnetic-tuning switch as a new kind of magnonics device.

\emph{Model}--The schematic diagram is illustrated in FIGs. 1(a) and (b).
We examine two cavity photon modes $\hat{c_1}$ and $\hat{c_2}$, interacting with two connectors $\hat{a_1}$ and $\hat{a_2}$, and the coupling strength denoted as $\kappa_1$, $\kappa_2$, $\kappa_3$, $\kappa_4$, respectively. 
The anti-resonance mode $\hat{c_3}$ is produced by the combined influence of $\hat{c_1}$, $\hat{c_2}$, $\hat{a_1}$ and $\hat{a_2}$\cite{PhysRevApplied.22.064036}.
We present a YIG wafer to provide magnon mode $\hat{m}$, which couples with anti-resonance mode and cavity modes, and the coupling strength is denoted as $g_1$ and $g_2$ respectively. 

\begin{figure}[htbp]
\centering
\includegraphics[width=8.6 cm]{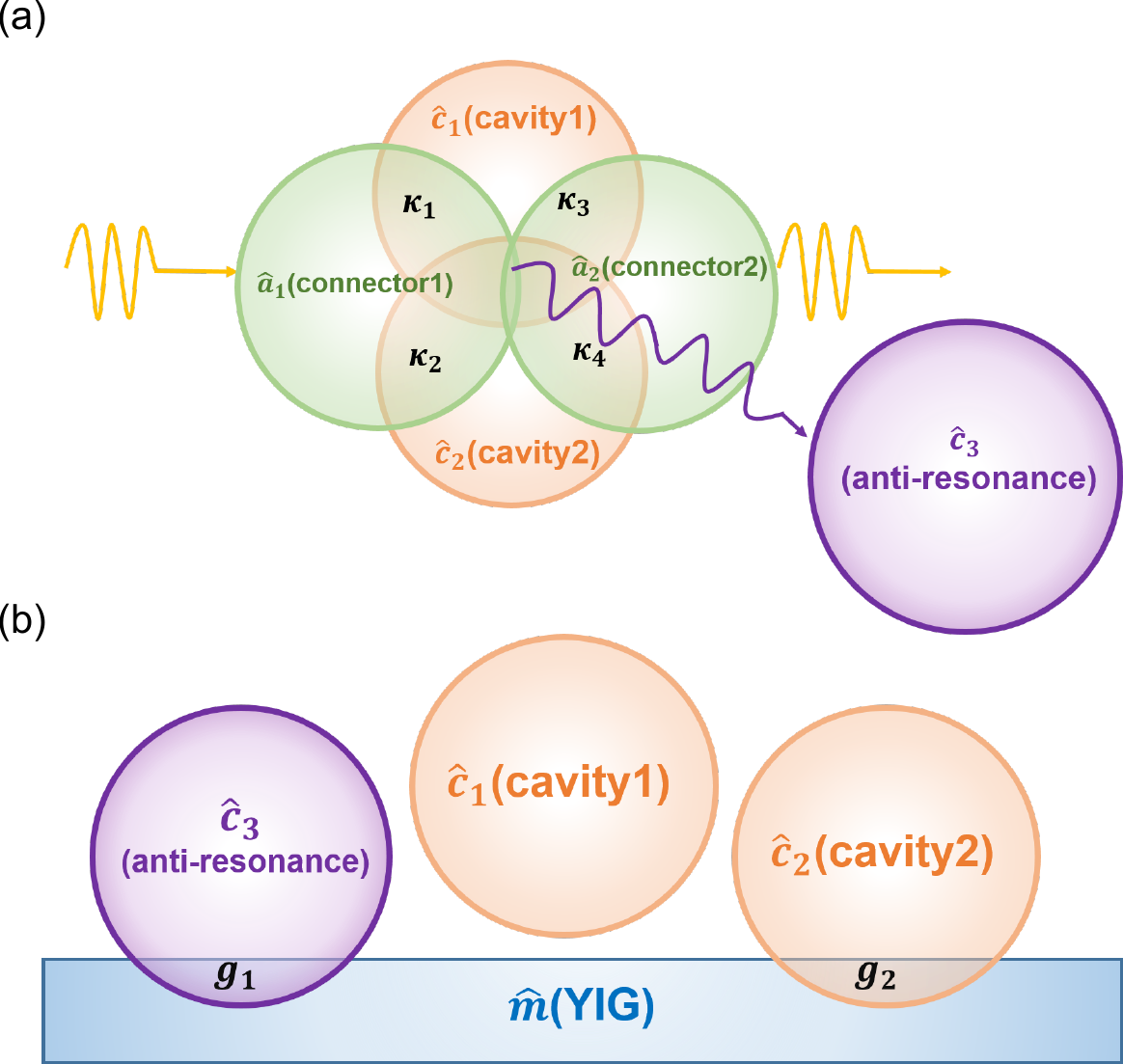}
\caption{(a) Schematic diagram of the anti-resonance generated by the joint action of connectors and cavity modes. (b) Schematic diagram of magnon-photon couplings in this work. $\hat{c_1}$ and $\hat{c_2}$ are the cavity modes. $\hat{c_3}$ is anti-resonance mode. $\hat{a_1}$ and $\hat{a_2}$ are the connector modes. $\hat{m}$ is the magnon mode.}
\label{Fig:1}
\end{figure}

\emph{Results and discussion}--FIG. 2(a) illustrates our experimental setup. 
The couplings are induced in a resonant cavity and a piece of YIG wafer (procured from HF-Kejing Material Technology Inc.). 
The dimensions of the rectangular cavity are as follows: 80 mm in length, 40 mm in width, and 5 mm in height.
The YIG wafer is 5.66 mm in diameter and 0.52 mm in thickness. 

Our experiment is conducted in the cavity depicted in FIG. 2(a), with the junction of the x, y, z axes designated as the origin of the coordinate system (0, 0, 0). We installed two connectors (bought from Nanjing Ningyue Communication Technology Inc.)\cite{supplementary} at the top of the rectangular cavity. The connectors extend into the cavity by 2.5 mm. The coordinates of the two connectors are (10 mm, 20 mm, 5 mm) and (70 mm, 20 mm, 5 mm), respectively. A Vector Network Analyzer (VNA) (keysight N5227B PNA) is employed to test the input and output of the signal, with port 1 and port 2 of the VNA linked to connector 1 and connector 2, respectively. FIG. 2(b) illustrates the amplitude and phase of the eigenmode of the rectangular cavity, with the red solid line denoting the experiment data and the blue dashed line denoting the simulation data. In the simulation of cavity eigenmodes, we incorporate the effects of two connectors. The cavity photon modes we are interested in are $\hat{c_1}$ and $\hat{c_2}$, with frequencies of 11.438 GHz and 11.76 GHz, respectively, and dissipation rates denoted as $\beta_1$ and $\beta_2$, respectively, as detailed in the supplementary material\cite{supplementary}. The magnon modes are generated by the YIG wafer, located at (20 mm, 15 mm, 0 mm). A static magnetic field $H$ is applied along the z direction, effectively exciting the magnon modes in the YIG wafer\cite{J.Phys.D:Appl.Phys.50.205003}; with the magnon dissipation rate of a single crystal YIG wafer quantified as $\alpha=3\times 10^{-5}$\cite{Phys.Rep.229.81}. According to the schematic of FIG. 1(a), $\hat{c_1}$, $\hat{c_2}$ interact with $\hat{a_1}$, $\hat{a_2}$ to generate an anti-resonance mode $\hat{c_3}$, characterized by a frequency of 11.367 GHz and a dissipation rate of $\beta_3$\cite{supplementary}. FIG. 2(c) illustrates the amplitude and phase of these three modes, with the black solid line denoting the amplitude and the blue solid line denoting the phase. The phase of $\hat{c_1}$ transition from high to low, termed a ``negative'' phase change, whereas the phase of $\hat{c_2}$ and $\hat{c_3}$ transition from low to high, referred to as a ``positive'' phase change. Given that the phase changes of $\hat{c_2}$ and $\hat{c_3}$ are positive, in contrast to $\hat{c_1}$, the coupling of the YIG wafer with $\hat{c_3}$ is characterized as dissipative coupling, whereas there is no coupling with $\hat{c_1}$, and the coupling with $\hat{c_2}$ is characterized as coherent coupling\cite{PhysRevApplied.22.064036,supplementary}.

\begin{figure}[htbp]
\centering
\includegraphics[width=8.6 cm]{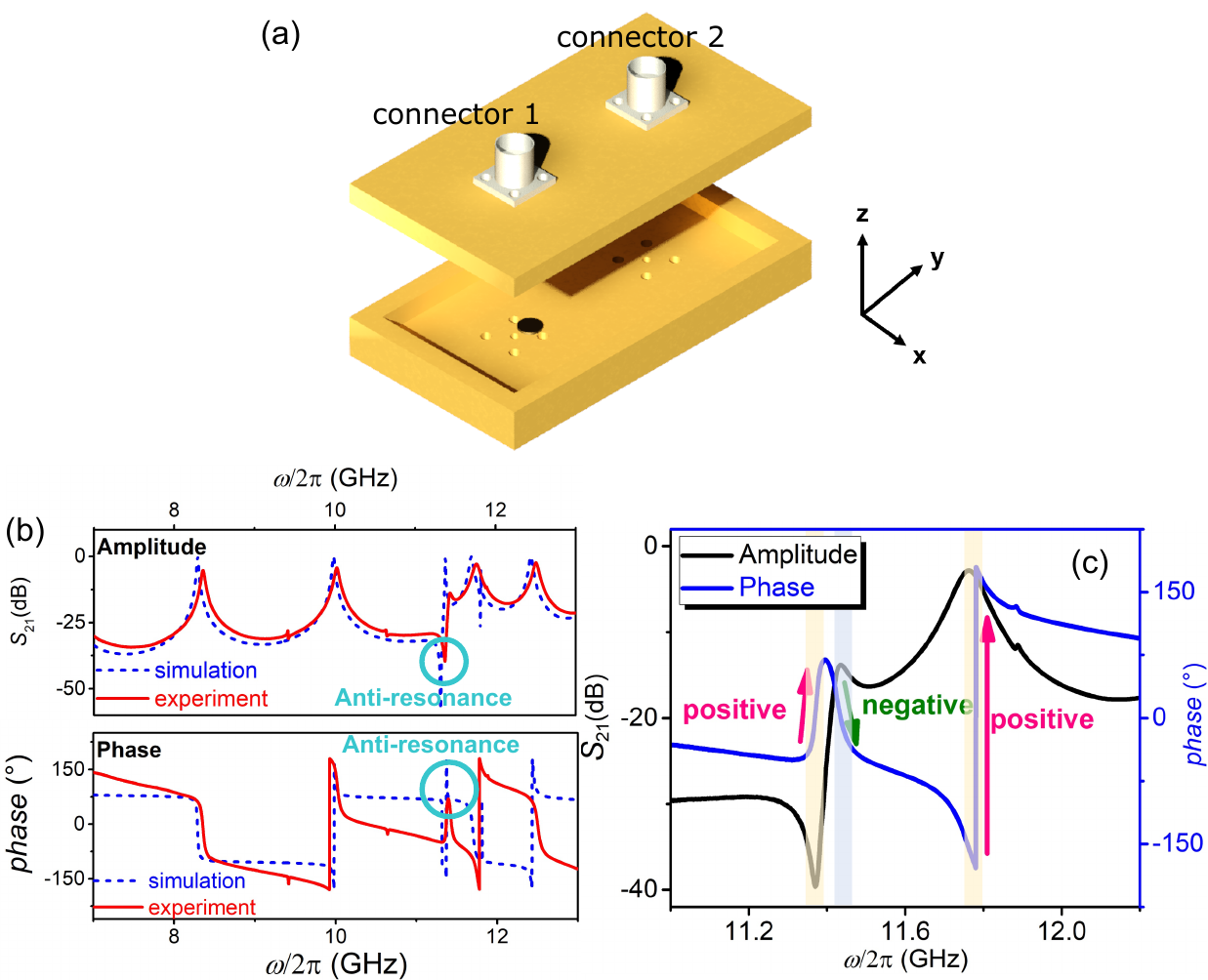}
\caption{(a) Schematic diagram of the experimental device (rectangular cavity). The position of the YIG wafer is (20 mm, 15 mm, 0 mm). The spectrum is measured using VNA through connector 1 and connector 2. (b) The amplitude and phase of of the rectangular cavity without applied magnetic field from 7 GHz to 13 GHz. The red solid line and the blue dashed line stands for the experiment data and the simulation data, respectively. (c) Experimental data of amplitude (the black solid line) and phase (the blue solid line) from 11 GHz to 12.2 GHz.}
\label{Fig:2}
\end{figure}

By varying the applied magnetic field $H$ and measuring the transmission coefficients, we obtain FIG. 3(a). The FMR mode (yellow dashed line) is described by the Kittel equation\cite{PhysRev.73.155}

\begin{equation}\label{1}
 f_K=\gamma\sqrt{(H+(N_x-N_z)M_s)(H+(N_y-N_z)M_s)},
\end{equation}
where the gyromagnetic ratio $\gamma$ is 2.68 MHz/Oe, the saturation magnetization $M_s$ of the YIG wafer is 1750 G, and $H$ is the applied magnetic field. The demagnetization of the wafer is non-uniform, rendering it impossible to obtain analytically. We utilize the oblate ellipsoid demagnetization factor from the textbook\cite{Physics.of.Ferromagnetism}. The demagnetization factor $N_x$, $N_y$ in the x and y direction is 0.07, and the demagnetization factor $N_z$ in the z direction is 0.86. FIG. 3(a) illustrates the generation of other higher-order spin wave modes in addition to the FMR mode. Given that the external magnetic field is oriented along the z direction, we ascertain that these higher-order spin wave modes are classified as FVMSW (cyan dashed line)\cite{J.Phys.D:Appl.Phys.50.205003,J.Phys.D:Appl.Phys.43.264002,J.Phys.Chem.Solids.19.308}, with FVMSW described by the subsequent formula\cite{J.Phys.D:Appl.Phys.43.264002}:

\begin{equation}\label{2}
 f_\text{FVMSW} = \sqrt{f_K \left( f_K + f_M \left( 1 - \frac{1 - \exp(-k d_0)}{k d_0} \right) \right)},
\end{equation}
where $f_K$ is the FMR mode, $f_M=\gamma M_s$, $d_0$ is the thickness of the YIG wafer with a value of 0.52 mm, and $k$ is the wave vector. The Bessel function of the first kind is introduced to obtain the values of the wave vector. $\mu_{nm}$ stands for the $m$th eigenvalue of the Bessel function of the first kind of order n. In our experiment, the wave factor could be defined as $k=2\mu_{nm} /r$. $r$ is the radius of the YIG wafer with a value of 2.83 mm. The value of $\mu_{nm}$ can be found from Ref.\cite{J.Appl.Mech.32.239}: $\mu_{01}$=2.405, $\mu_{11}$=3.832,  $\mu_{21}$=5.136, $\mu_{31}$=6.380, $\mu_{41}$=7.588, $\mu_{51}$=8.772. Then we put these values into Eq. (2) and obtained FVMSW modes, as shown in Fig. 3(a) (marked cyan dashed lines). The modes with a frequency lower than that of the FMR mode are the non-uniform magnetization modes. Attractively, the couplings induced by different magnon modes are distinctly different; the coupling between the FMR mode and the anti-resonance mode is characterized as dissipative coupling. Conversely, the coupling of the FVMSW modes and the anti-resonance mode is characterized as coherent coupling.

\begin{figure*}[htbp]
\centering
\includegraphics[width= 17.2 cm]{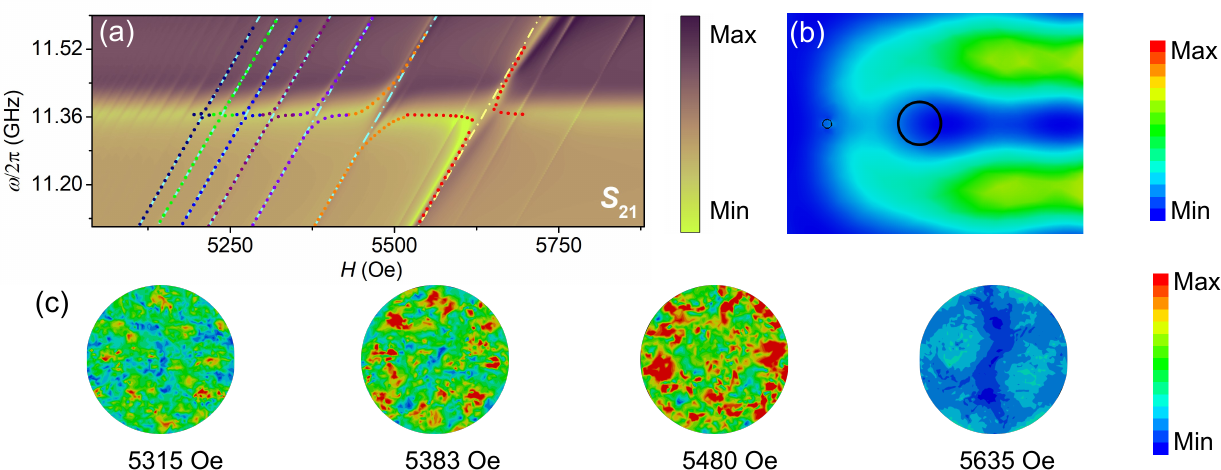}
\caption{(a) The density mapping image of the amplitude of the transmission coefficients through the cavity as a function of frequency and the applied static magnetic field. 
The deeper colour of the image expresses the larger microwave transmission loss. $S_{21}$ mapping at the magnetic field applied along the $z$ direction.
(b) Simulation of the microwave magnetic field distribution at 11.367 GHz in the cavity system at an applied magnetic field of 0, and the black circle is the position of the YIG wafer.
(c) Simulation of the microwave magnetic field distribution of representative magnon modes at 11.367 GHz.
The corresponding magnetic fields are 5315 Oe ($\mu_{21}$), 5383 Oe ($\mu_{11}$), 5480 Oe ($\mu_{01}$) and 5635 Oe (FMR).}
\label{Fig:3}
\end{figure*}

 To describe our experimental system, we give out the non-Hermitian Hamiltonian\cite{PhysRevB.98.024406}:

 \begin{equation}\label{3}
\frac{\hat{H}}{\hbar} = \tilde{\omega}_c\hat{a}^\dagger \hat{a} + \tilde{\omega}_m\hat{b}^\dagger \hat{b} + \left( J - i\Gamma e^{i\theta} \right)\left( \hat{a}^\dagger \hat{b} + \hat{b}^\dagger \hat{a} \right),
\end{equation}
where $\tilde{\omega}_c = \omega_c - i \beta_3 $, $\tilde{\omega}_m = \omega_m - i \alpha$, $\hat{a}$ ($\hat{a}^\dagger$) and $\hat{b}$ ($\hat{b}^\dagger$) are operators of antiresonance photon and magnon annihilation (creation), respectively. Here, $\alpha$ and $\beta_3$ represent the dissipation rates of the magnon mode and the anti-resonance mode, respectively. The phase difference $\theta$ of the loaded microwave at port 1 and port 2 of VNA is 0 and $\pi$, $J$ denotes the coupling strength of the coherent coupling, and $\Gamma$ denotes the coupling strength of the dissipative coupling. Eq.(3) can be solved in the standard way to obtain two eigenvalues:

\begin{widetext}
\begin{equation}\label{4}
\begin{aligned}
\tilde{\omega}_{\pm} &= \frac{1}{2} \left[ \omega_c + \omega_m - i (\beta_3 + \alpha) \pm \sqrt{\left[ (\omega_c - \omega_m) - i (\beta_3 - \alpha) \right]^2 + 4 \left( J - i e^{i \theta} \Gamma \right)^2  } \right].
\end{aligned}
\end{equation}
\end{widetext}

In order to understand the experimental phenomena more distinctly, we individually fit these couplings separately. 
The coupling between the anti-resonance mode and FMR mode is represented by the red dots in FIG. 3(a).
It is determined that $\Gamma/2\pi$=70 MHz and $J/2\pi$=1 MHz, sindicating a coupling mostly governed by dissipative coupling.
Subsequently, we also fit the couplings between the anti-resonance mode and the FVMSW modes.
When $\mu_{10}$=2.405, the coupling between the anti-resonance mode and FVMSW mode is represented with orange dots, $\Gamma/2\pi$=5 MHz, $J/2\pi$=40 MHz, indicating a coupling mostly governed by coherent coupling.
Similarly, the couplings induced by the anti-resonance mode with different FVMSW modes are represented by violet, purple, blue, green, and navy dots.
The coupling strength and coupling modes are detailed in Table 1. 
Obviously, the anti-resonance mode is dissipatively coupled to the FMR mode and coherently coupled to the FVMSW modes. 
FIG. 3(b) illustrates the microwave magnetic field intensity distribution within the cavity at the frequency of 11.367 GHz.
The black circle stands for the position of the YIG wafer.
The strength of the microwave magnetic field at the position of the YIG wafer is notably weak.
We employed the finite element method to compute the spin wave mode in our experiment at 11.367 GHz, as illustrated in FIG. 3(c).
The rightmost one displays the microwave magnetic field distribution of the YIG wafer at 5635 Oe, corresponding to the FMR mode at the frequency of 11.367 GHz in FIG. 3(a). 
The microwave magnetic field distribution at dissipative couplings of 5635 Oe is notably weak, consistent with the microwave magnetic field distribution observed in the cavity in FIG. 3(b).
The rest of three graphs in FIG. 3(c) represent the microwave field distribution of the YIG wafer at coherent couplings of 5315 Oe ($\mu_{21}$), 5383 Oe ($\mu_{11}$) and 5480 Oe ($\mu_{01}$), respectively.
The microwave magnetic field distribution in the YIG wafer under these three fields is notably strong, contrasting sharply with the microwave magnetic field distribution in the cavity depicted in FIG. 3(b).
Consequently, we deduce that when the microwave field distribution of the YIG wafer aligns with its position in the cavity, the spin wave exhibits dissipative coupling with the anti-resonance mode.
Otherwise, it constitutes coherent coupling. It is important to note that dissipative couplings can only take place when interacting with the anti-resonance mode of the cavity, the resonance modes are all coherent couplings.

\begin{figure}[htbp]
\centering
\includegraphics[width=8.6 cm]{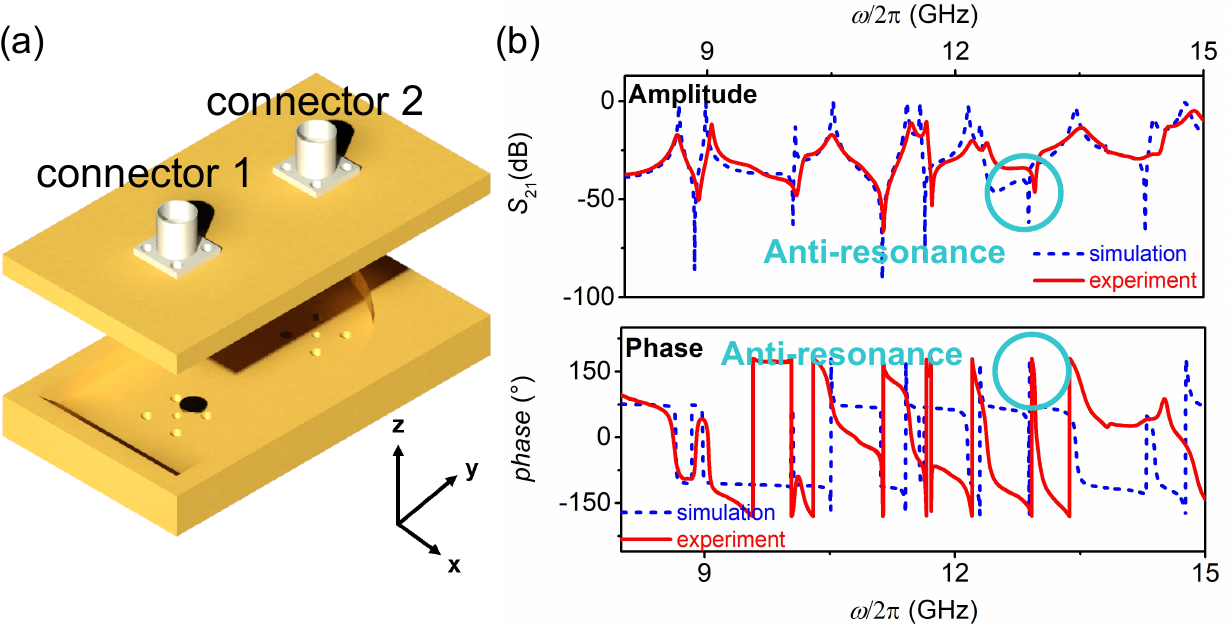}
\caption{(a) Schematic diagram of the experimental device (quadrant-stadium cavity). The position of the YIG wafer is (20 mm, 15 mm, 0 mm). The spectrum is measured using VNA through connector 1 and connector 2. (b) The amplitude and phase of of the rectangular cavity without applied magnetic field from 8 GHz to 15 GHz. The red solid line and the blue dashed line stands for the experiment data and the simulation data, respectively.}
\label{Fig:4}
\end{figure}

In the description of classical dynamics, the rectangular cavity is an integrable system whose microwave electromagnetic field pattern follows the Poisson distribution. However, for chaotic systems, the microwave electromagnetic field pattern obeys Gaussian distribution, which makes the electromagnetic field distribution of the system tend to be chaotic\cite{Deterministicchaos,QuantumSignaturesofChaos}. To prevent our conclusion from being influenced by the different systems, we set up a chaotic system with a quadrant-stadium cavity\cite{PhysRevLett.103.064101,PhysRevE.81.036205,supplementary}, as seen in FIG. 4(a). FIG. 4(b) illustrates the amplitude and phase of the cavity eigenmode, with the red solid line denoting the experiment data and the blue dashed line denoting the simulation data. We focus on the anti-resonance at the frequency of 13.05 GHz, and the dissipation rate is denoted as $\beta_6$\cite{supplementary}. The coordinate of the YIG wafer is (20 mm, 15 mm, 0 mm).

\begin{figure*}[htbp]
\centering
\includegraphics[width=17.2 cm]{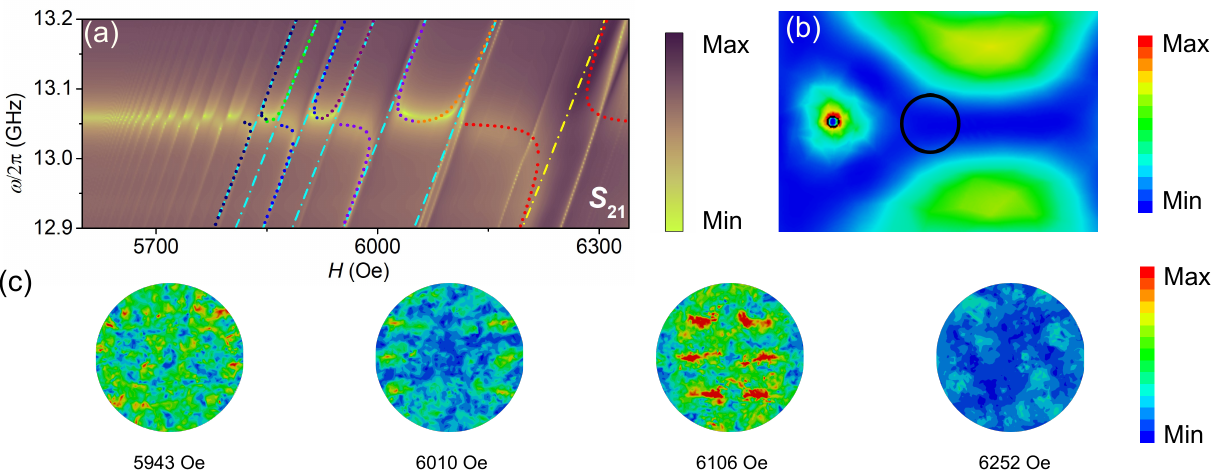}
\caption{ (a) The density mapping image of the amplitude of the transmission coefficients through the cavity as a function of frequency and the applied static magnetic field. The deeper color of the image expresses the larger microwave transmission loss. $S_{21}$ mapping at the magnetic field added along direction $z$. (b) Simulation of the microwave magnetic field distribution at 13.05 GHz at an applied magnetic field of 0, and the black circle is the position of the YIG wafer. (c) Simulation of the microwave magnetic field distribution of representative magnon modes at 13.05 GHz. The corresponding magnetic fields are 5943 Oe ($\mu_{21}$), 6010 Oe ($\mu_{11}$), 6106 Oe ($\mu_{01}$) and 6252 Oe (FMR).}
\label{Fig:5}
\end{figure*}

\begin{table*}[htbp]
\caption{Coupling strength and coupling mode}
\label{tab_demo}
\resizebox{0.8\linewidth}{!}{
\scriptsize 
\begin{tabular}{ccccccccccccccc}
\toprule
\multicolumn{15}{c}{\textbf{Rectangular Cavity}} \\ 
\midrule
\textbf{Coupling} & \multicolumn{2}{c}{red} & \multicolumn{2}{c}{orange} & \multicolumn{2}{c}{violet} & \multicolumn{2}{c}{purple} & \multicolumn{2}{c}{blue} & \multicolumn{2}{c}{green} & \multicolumn{2}{c}{navy} \\
\cmidrule(r){2-3} \cmidrule(r){4-5} \cmidrule(r){6-7} \cmidrule(r){8-9} \cmidrule(r){10-11} \cmidrule(r){12-13} \cmidrule(r){14-15}
\multirow[b]{2}{*}{\textbf{Strength (MHz)}} & $J$ & $\Gamma$ & $J$ & $\Gamma$ & $J$ & $\Gamma$ & $J$ & $\Gamma$ & $J$ & $\Gamma$ & $J$ & $\Gamma$ & $J$ & $\Gamma$ \\
 & 1 & 70 & 40 & 5 & 20 & 5 & 10 & 5 & 8 & 2 & 6 & 1 & 5 & 1 \\
\midrule
\textbf{Mode} & \multicolumn{2}{c}{dissipative} & \multicolumn{2}{c}{coherent} & \multicolumn{2}{c}{coherent} & \multicolumn{2}{c}{coherent} & \multicolumn{2}{c}{coherent} & \multicolumn{2}{c}{coherent} & \multicolumn{2}{c}{coherent} \\
\hline
\hline 
\multicolumn{15}{c}{\textbf{Quadrant-stadium Cavity}} \\ 
\midrule
\textbf{Coupling} & \multicolumn{2}{c}{red} & \multicolumn{2}{c}{orange} & \multicolumn{2}{c}{violet} & \multicolumn{2}{c}{purple} & \multicolumn{2}{c}{blue} & \multicolumn{2}{c}{green} & \multicolumn{2}{c}{navy} \\
\cmidrule(r){2-3} \cmidrule(r){4-5} \cmidrule(r){6-7} \cmidrule(r){8-9} \cmidrule(r){10-11} \cmidrule(r){12-13} \cmidrule(r){14-15}
\multirow[b]{2}{*}{\textbf{Strength (MHz)}} & $J$ & $\Gamma$ & $J$ & $\Gamma$ & $J$ & $\Gamma$ & $J$ & $\Gamma$ & $J$ & $\Gamma$ & $J$ & $\Gamma$ & $J$ & $\Gamma$ \\
 & 5 & 90 & 30 & 4 & 4 & 30 & 20 & 2 & 2 & 20 & 10 & 1 & 7 & 9.8 \\
\midrule
\textbf{Mode} & \multicolumn{2}{c}{dissipative} & \multicolumn{2}{c}{coherent} & \multicolumn{2}{c}{dissipative} & \multicolumn{2}{c}{coherent} & \multicolumn{2}{c}{dissipative} & \multicolumn{2}{c}{coherent} & \multicolumn{2}{c}{dissipative} \\
\bottomrule
\end{tabular}
}
\end{table*}

By varying the applied magnetic field $H$ and measuring the transmission coefficients, we obtain FIG. 5(a).
In a similar manner, we individually fit these couplings separately. The FMR mode (yellow dashed line) is described by the Eq.(1).
The value of $\mu_{nm}$ can be found from Ref.\cite{J.Appl.Mech.32.239}: $\mu_{01}$=2.405, $\mu_{11}$=3.832,  $\mu_{21}$=5.136, $\mu_{31}$=6.380, $\mu_{41}$=7.588, $\mu_{51}$=8.772. Then we put these values into Eq. (2) and obtained FVMSW modes, as shown in Fig. 5(a) (cyan dashed lines). 
Attractively, the coupling between the anti-resonance mode and the spin wave modes occur alternately with the coherent coupling and the dissipative coupling.
We fit the coupling using Eq.(4), represented by red, orange, violet, purple, blue, green, and navy dots.
The coupling strength and coupling mode are presented in Table 1.
Using the finite element method, the microwave magnetic field intensity distribution of the cavity at 13.05 GHz is depicted in FIG. 5(b).
The black circle stands for the position of the YIG wafer, and the strength of the microwave magnetic field at the position of the YIG wafer is notably weak.
The microwave magnetic field distribution of representative magnon modes at 13.05 GHz are computed, as illustrated in FIG. 5(c).
The four diagrams delineate the magnetic field distribution of the YIG wafer at 5943 Oe ($\mu_{21}$), 6010 Oe ($\mu_{11}$), 6106 Oe ($\mu_{01}$) and 6252 Oe (FMR).
The microwave magnetic field distributions of 6016 Oe and 6252 Oe are notably weak, consistent with the magnetic field distribution observed in the cavity in FIG.5(b); thus, the associated red and violet dots represent dissipative coupling. 
Conversely, the microwave magnetic field distribution in the YIG wafer under 5943 Oe and 6106 Oe are notably strong, contrasting sharply with the magnetic field distribution in the cavity depicted in FIG. 5(b); thus, the associated origin and purple dots indicate coherent coupling.
The outcome is identical to that in the rectangular cavity, proving an inherent characteristic.

\begin{figure}[htbp]
\centering
\includegraphics[width=8.6 cm]{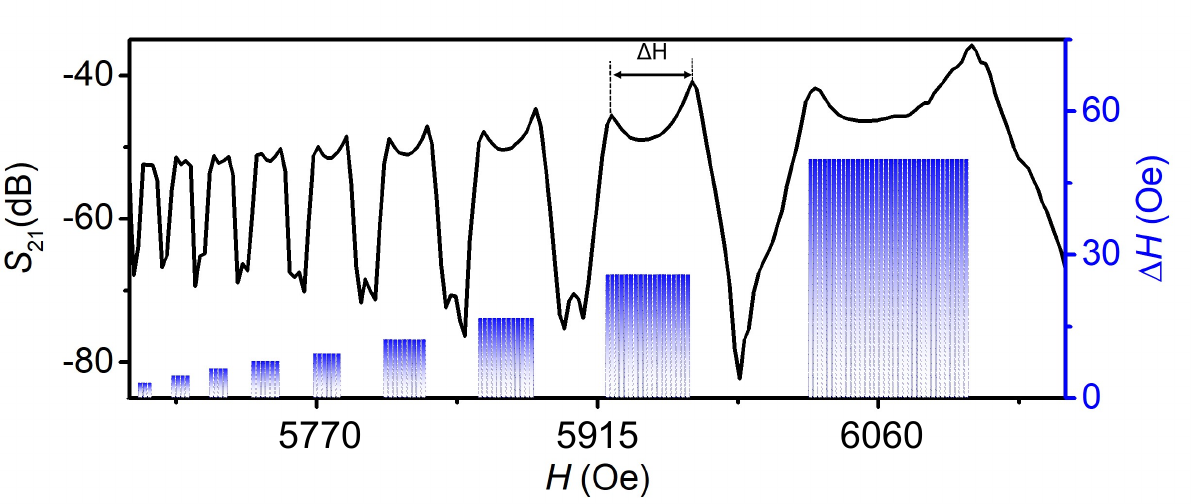}
\caption{The black solid line represents the change of the transmission parameter with the applied magnetic field at the anti-resonance frequency of 13.05 GHz. The blue histogram stands for the variation trend of $\Delta H$.}
\label{Fig:6}
\end{figure}

Such unique behaviors dynamics enable flexible controllability. In addition to the alternate dissipative coupling and coherent coupling that we have shown, the lower polarized mode ($\omega_-$) of the coherent couplings tend to vanish almost due to the interference of coherent and dissipative couplings\cite{PhysRevLett.123.127202}. 
This phenomenon induces the continuous transformation of the transmission signal at the anti-resonance frequency of 13.05 GHz in the coupling system. 
For clarity, we only plot the transmission coefficient at 13.05 GHz from Fig. 5(a), illustrated in Fig. 6. The black line illustrates the variation in the transmission coefficient of system with respect to the applied magnetic field at a frequency of 13.05 GHz. As depicted in the figure, it is evident that the coupling between each magnon mode and anti-resonance mode enhances signal transmission within a specific range of magnetic fields $\Delta$$H$. Additionally, As the order of the magnon mode decreases, $\Delta$$H$ becomes larger, described by histogram. As we know, anti-resonance is a phenomenon in resonant systems where the system exhibits a significant suppression of response, effectively blocking the transmission of energy at specific frequencies. This indicates that continuous modulation of both coherent and dissipative couplings enables selective transmission of microwave signals through specific range of the applied magnetic field within our system. It also paves a new way to design a continuously magnetic-tuning switch.

 \emph{Conclusion.}-- In this work, two distinct kinds of couplings, namely the dissipative and coherent coupling are realized synchronously by the interaction between multiple different magnon modes and a same anti-resonance mode in a large range of the applied magnetic field. 
We demonstrate that in a quasi-close cavity, dissipative coupling could only appear at the frequency of the anti-resonance mode and distribution of the microwave magnetic field of the magnon mode at this frequency needs to be as weak as possible. 
Besides, this law is experimentally realized in both classical and chaotic cavities. 
By controlling the dissipative and coherent coupling alternately, the lower polarized mode of the coherent coupling tends to disappear almost due to the competition between dissipation and coherent coupling. 
It allows microwave signals at the frequency of anti-resonance mode to be selectively transmitted over a large applied magnetic field range.
It is reasonable to believe that the quality of the passable microwave signals can be improved by improving the quality of the couplings. 
It may promote the construction of a new kind of magnonics device like magnetic-tuning switch.

The authors would like to thank Prof. Peng Yan, Prof. Jinwei Rao, for useful discussions and suggestions.
This work is supported by the National Natural Science Foundation of China (NSFC) (Nos. 52471200, 12174165 and 52201219).

\bibliography{Ref.bib}
\end{document}